
%
%
\magnification=1200\overfullrule=0pt\baselineskip=15pt
\hoffset=.65cm
\vsize=22truecm \hsize=15truecm \overfullrule=0pt\pageno=0
\font\titlefont=cmbx12\font\sectnfont=cmbx10
\def\mname{\ifcase\month\or January \or February \or March \or April
           \or May \or June \or July \or August \or September
           \or October \or November \or December \fi}
\def\date{\hbox{\strut\mname \number\year}}
\def\tifrnum{\hbox{TIFR/TH/92-24\strut}}
\def\banner{\hfill\hbox{\vbox{\offinterlineskip\tifrnum}}\relax}
\def\manner{\hbox{\vbox{\offinterlineskip\tifrnum\date}}
               \hfill\relax}
\footline={\ifnum\pageno=0\manner\else\hfil\number\pageno\hfil\fi}
%
\banner\bigskip\begingroup\titlefont\obeylines\noindent
Can the Quark-Gluon Plasma in the Early Universe
\hfil be supercooled?\hfil
\endgroup\bigskip\bigskip
\centerline{B.~Banerjee and R.~V.~Gavai}
\centerline{Theoretical Physics Group}
\centerline{Tata Institute of Fundamental Research}
\centerline{Homi Bhabha Road}
\centerline{BOMBAY 400 005, INDIA}
\
\bigskip\bigskip\centerline{{\sectnfont ABSTRACT}}\bigskip
The quark-hadron phase transition in the early universe can produce
inhomogeneities in the distribution of nucleons, which in turn
affect the primordial nucleosynthesis. In all the investigations
of this problem it has been assumed that the degree of supercooling
of the quark-gluon plasma after the phase transition is large
enough to produce a significant rate of nucleation of hadrons.
Using the latest results of finite temperature lattice QCD and
the finite size scaling theory, we argue that the degree of
supercooling is in fact extremely small and hence the nucleation
rate is negligible.

\vfil\eject


The implications of the transition from the deconfined quark-gluon
plasma phase to the confined hadron phase in the early universe for
the primordial nucleosynthesis have been studied extensively in recent years
[1-11]. The following phase separation scenario
has been used in these investigations.
In the early universe when the temperature is high the unconfined quarks and
 gluons form a plasma. As the universe cools due to expansion and
reaches the phase transition temperature, $T_c$, the confined hadron
phase does not appear immediately since the phase transition is
 assumed to be of first order. The universe has to cool somewhat
below~$T_c$~before the probability for nucleating bubbles becomes
significant. The supercooled phase is metastable and decays
into hadronic bubbles.

The rate of nucleation of hadron bubbles has been calculated
 [4,12] using the theory of homogeneous nucleation [13]. In
this theory the new phase is assumed to arise through spontaneous
fluctuations in the supercooled phase. The nucleation rate per unit volume,
 $p(T)$, at a temperature, $T$, is given by
$$ p(\eta) \simeq CT_c^4 \exp (-{16\pi\over 3}{\sigma^3
\over{T_c L^2\eta^2}})
\eqno(1)$$
where~$C$~is a constant of order unity, $\sigma$, the surface
tension of the hadronic bubble, $L$, the latent heat per unit
volume of the transition and $\eta$, the degree of supercooling,
$\eta=(T_c-T)/T_c$.  As the supercooling is assumed to be small, after a
hadron bubble has been nucleated it grows explosively as a deflagration bubble
[12,14] and drives a weak shock-wave which expands into the surrounding
quark phase with a velocity just above sound speed~$v_s\simeq 1/\sqrt3$~.
The plasma gets reheated by the shock-wave and further nucleation is
inhibited due to the steep fall-off in Eq.\ (1).
The fraction of the universe affected by the shock front
preceding the growing bubble is $$ g(t)= \int_{t_c}^t
p(t^\prime){4\pi\over3}\,v_s^3(t-t^\prime)^3
\, dt^\prime~~,~~
\eqno(2)$$
where $t_c$ is the time when the universe cools through $T_c$.
Here the temperature $T$ in $p(T)$ has been replaced by time
$t$ using the  Einstein relation between $T$ and the age of
the universe,$t$,
$$ t= {3 \over {2\pi(41\pi)^{1/2}}}{M_{Pl}\over T^2}
\eqno(3)$$
In calculating $t$ the quark degrees of freedom have been
included\footnote{${}^\dagger$}{We have used here the ideal gas expression
for the energy density. While it is now known[15] that near $T_c$ there
are strong non-perturbative effects, they will affect Eq.\ (3) by a factor
of O(1) and have therefore been ignored.}.  The total number density
of nucleation sites is then
$$ N_n = \int_{t_c}^\infty f(t)p(t)dt~~,~~ \eqno(4)$$
where
$$f(t)=1-g(t)~~.~~ \eqno (5)$$
Since the supercooling, that is, $\eta$ is
small, $p(t)$ rises rapidly with $t$ and therefore $f(t)$ behaves almost
like a step function~$\theta(t_h-t)$~[4, 12], where $t_h$ is the time when
the universe has been reheated.  $N_n$ can then  be approximated as
$$ N_n = \int_{t_c}^{t_h} p(t)\, dt~~.~~ \eqno(6)$$
By expanding $p(t)$ around $t_h$ one obtains [4],
$$ p(t)= p(t_h)\exp(-\alpha(t_h-t)) ~~,~~\eqno(7)$$
where
$$\alpha = 1.6528\times10^{-7}{\gamma^2\over\eta_h^3}{\left(T_c
\over {MeV}\right)}^2 m^{-1}~~,~~
\eqno(8)$$
$\gamma^2$ is a dimensionless combination of $\sigma$, $L$, and $T_c$,
$$ \gamma^2={\sigma^3 \over{L^2 T_c}}
\eqno(9) $$
and $\eta_h$ can be calculated from the condition $f(t_h)
=0$, i.e.,
$$\int_{t_c}^{t_h}~dt~ p(t^\prime){4\pi\over3}v_s^3(t-t^\prime)=1 ~~.~~
\eqno(10)$$
Substituting Eq.\ (7) in Eq.\ (10) and carrying out the integration
one finds the equation determining $\eta_h$ to be,
$$ x^6\exp(x) = 2.21\times 10^{-2}\left({M_{Pl}\, \gamma^2}\over
T_c\right)^4 ~~,~~\eqno (11)$$
where $$x = {16\pi\over3} {\gamma^2\over\eta_h^2}$$
(We have put $C=1$.).
An approximate solution of Eq.\ (11) given in ref.\ 10 is
$\eta_h= 0.4\gamma$.  Note that the RHS of Eq.\ (11) is totally dominated
by the very large ratio of the Planck scale and the QCD scale, which is
typically $\sim 10^{19}$. Thus, a large variation of $\gamma$, over several
orders of magnitude but still sufficiently small compared to the ratio,
changes the solution very little, making it almost independent of the
details of the QCD thermodynamics.

One of the parameters important for primordial nucleosynthesis
is the mean separation per nucleation site, $l\simeq N_n^{-1/3}$,
which can be obtained from Eq.\ (6) by evaluating the integral
using Eqs.\ (7)-(8) and the solution to Eq.\ (11) given above.
One obtains
$$ l = {(8\pi)^{1/3}}{v_s\over \alpha}~~.~~$$
A more detailed calculation by Meyer et al [10] gives
$$l = 4.38{v_s\over\alpha}~~.~~ \eqno(12)$$

During the subsequent evolution of the universe, when the hadronization
of the quark-gluon plasma is complete, the mean free paths of
neutrons and protons in the surrounding medium of hadrons, electrons,
 photons and neutrinos become different leading to a segregation
such that high-density proton-rich and low-density neutron-rich
regions develop. Nucleosynthesis then takes place among these
inhomogeneously distributed nucleons [3]. It is clear that for
this picture of nucleosynthesis to be valid $l$ should be
greater than the comoving diffusion length of the protons, which is
$\sim0.5$~m. The fluctuations would otherwise damp out entirely before
the onset of nucleosynthesis.

It is apparent from Eq.\ (8), (9) and (12) that the degree of
supercooling
and hence~$l$~depend only on the bulk thermodynamic properties,
namely, $v_s$, $\sigma$, $L$ and $T_c$ of the strongly interacting
matter. One knows, however, from finite size scaling theory [16]
that the degree of metastability is also decided by the size of the
system relative to its correlation length, $\xi(T_c)$ at the
transition point. Indeed , the width, $\Delta T$ of the metastable
region for a first order phase transition goes to zero as the
volume, V, increases:
$$ {\Delta T\over T_c}\equiv {T\over T_c}- 1= {A\over V}, \eqno(13)$$
where A is a constant of~$O(1)$~if V is measured in units of
$\xi(T_c)$.

In order to proceed further and see the consequences of Eq.\ (13)
we shall assume that the world of quenched quantum chromodynamics
 (QCD) is a good approximation to ours. In particular, we wish to
use the precise information available from lattice QCD in this
approximation. Furthermore, we shall assume that asymptotic scaling
is valid so that we can relate the lattice results to the continuum
theory. We shall return to the discussion of the effects of these
assumptions later. The best estimates from lattice QCD for the
quantities required are as follows [15, 17, 18]:
$$\eqalign{ L/T_c^4 &= 1.8\pm 0.1 \cr
            \sigma/T_c^3 &= 0.024\pm 0.004 \cr
            T_c &= 235\pm 42~{\rm MeV}\cr}~~.~~ \eqno (14)$$
Note that the transition temperature is much larger here than the
value used in Ref.\ 4 from the MIT bag model. The other parameters
are roughly the same. In fact, using
$T_c$ as a scale, one sees that it is really the latent heat,$L$, which is
much smaller here ( by almost a factor of 8), while the surface tension
is within the range used in Ref.\ 4. Using these values in Eqs.(9) and
(12), one obtains $\gamma=2.066 \times 10^{-3}$ and
$l = 2.84\pm 1.25$~cm, which is much less than what is required even
if we ignore the question of the size of the universe at $T~=~T_c$.

If we now ask how much supercooling we can expect by taking size of the
universe into account the situation becomes considerably worse. We know
from lattice QCD [19] that $i)$ the physical correlation length in the
deconfined phase is $\xi(T_c)T_c = 1.38\pm 0.24$ and $ii)$ the width of
the metastable region in the QCD-coupling space for a system of
volume $V$ is given by
$$
\eqalign{\Delta \beta &= 0.29{\left(\xi^3\over V\right)}
^{0.984\pm 0.039},\cr &= 0.75{\left(VT_c^3\right)}^{-0.984\pm 0.039
}\cr}~~.~~
\eqno (15)$$
Further, by assuming the validity of the asymptotic scaling relation, the
width in the coupling $\beta$ can be related to the width of the
metastable region in temperature: $\Delta T/T_c = 1.122 \Delta \beta$.
Using the expression for the age of the universe, Eq.\ (3), the volume of
the universe at temperature $T_c$ is found to be
$$ VT_c^3 = 7.1\times 10^{55}{\left(200 MeV\over T_c\right)^3}~~.~~
\eqno (16)$$
Substituting in Eq.\ (15), one obtains the degree of supercooling
possible in the early universe to be
$$ {\Delta T \over T_c} \equiv \eta = 9.28\times 10^{-56\pm 2.2}
{\left(T_c\over {200 MeV}\right)}^{2.952\pm 0.117}~~.~~ \eqno(17)$$
Thus for any value of $T_c$ in the range given in Eq.\ (14) the universe
practically never supercools at all. Even the miniscule amount
indicated by Eq.\ (17) will not be sufficient for any bubble nucleation,
 since the nucleation probability goes to zero as exp$\left(-1/
\eta^2\right)$ as~$\eta$~$\rightarrow 0$.

Let us now discuss whether our assumptions made earlier could have
led to the negative result. Two crucial inputs were the validity of the
asymptotic scaling relation and the finite size scaling relation Eq.\
(14).  Due to its better precision, we used above the lattice data
obtained on lattices with four points in the temporal direction. It is
expected [15] for quenched QCD that scaling may have set in for such
lattices but the asymptotic scaling is violated by $\sim 40$-50 \%,
causing at most a factor of 2 change in Eq.\ (17).  The systematic
error in the finite size scaling relation of Eq.\ (15) due to this is more
difficult to estimate but one does find similar results for larger
temporal lattices than those used in ref.\ 19, suggesting again changes by
factors of O(1).   Indeed, within the quenched QCD approximation it
appears difficult to get away from the fact that the size of the universe
at $T_c$ measured in the units of the physical correlation length is
essentially infinite, yielding an almost null result for the supercooling
parameter.

The real universe of course has dynamical fermions and one must
therefore ask what happens in the case of full QCD. Unfortunately,
in this case one has to worry about more input parameters in addition
to the  usual technical problem of putting fermions on the lattice.
Assuming the current state of the art results, at best a very
weak first order phase transition may be indicated for the realistic
two light flavours and one heavier flavour case [20]. No signs of
a first order transition were found on a lattice of linear
size $4/T_c$, suggesting a still larger correlation length at $T_c$,
if finite at all. A larger correlation length at $T_c$ will
increase the possibility of appreciable supercooling since a weaker phase
transition will presumably yield broader region of metastability.
Furthermore, a weaker phase transition will have lower latent heat and
surface tension.  These, in turn, may reduce $\gamma$ and therefore also
the required degree of supercooling, $\eta_h$, to achieve the desired mean
free path.  Thus, the scenario proposed in refs. \ [1-4] becomes more
feasible for a very weak first order phase transition.  However, it still
looks rather improbable that the supercooling necessary for hadronic
bubble creation will be achieved unless $\gamma$ for the full QCD is lower
by several orders of magnitude from its value for the quenched QCD used
above.  Whether future results from simulations of full QCD will still
allow a small probability for homogeneous nucleation thus remains to be seen.
The alternative and more likely possibility is heterogeneous nucleation
caused by `impurities' left over from an earlier epoch.

\bigskip\line{\sectnfont Acknowledgements. \hfil}

One of us (B.Banerjee) wishes to thank D.N.Schramm for useful discussions
and for the hospitality at the Department of Astronomy
and Astrophysics, University of Chicago. His visit was
supported by NSF grants INT 91-101171DS and INT 87-15411.

\vfill\eject
\centerline{\sectnfont {References}}\bigskip

\item{[1]} M. Crawford
 and D. N. Schramm, Nature 298 (1982) 538; E. Witten, Phys. Rev. D
30 (1984) 538.
\item{[2]} J. H. Applegate and C. J. Hogan, Phys. Rev. D 31 (1985) 3037.
\item{[3]} J. H. Applegate, C. J. Hogan and R. Scherrer, Phys. Rev. D
 35 (1987) 1151.
\item{[4]} G. M. Fuller, G. J. Mathews and C. R. Alcock, Phys. Rev.
 D 37 (1988) 1380.
\item{[5]} R. A. Malaney and W. A. Fowler, Astrophys. J. 333 (1988) 14.
\item{[6]} H. Kurki-Suonio and R. A. Matzner, Phys. Rev. D 39 (1989)
1046.
\item{[7]} N. Terasawa and K.Sato, Phys. Rev. D 39 (1989) 2893.
\item{[8]} H. Kurki-Suonio, R. A. Matzner, K.A. Olive and
 D.N. Schramm, Astrophys. J. 353 (1990) 406.
\item{[9]} G. J. Mathews, B. S. Meyer, C. R. Alcock and G. M. Fuller,
Astrophys. J. 358 (1990) 36.
\item{[10]} B.S. Meyer, C. R. Alcock, G. J. Mathews and
G. M. Fuller, Phys. Rev. D 43 (1991) 1079.
\item{[11]} H. Kurki-Suonio, M. R. Aufderheide, F. Graziani,
 G. J. Mathews, B. Banerjee, S. M. Chitre and D. N. Schramm,
 Phys. Lett.B (submitted).
\item{[12]} K. Kajantie and H. Kurki-Suonio, Phys. Rev. D 34 (1986)
 1719.
\item{[13]} L. Landau and E. M. Lifshitz, {\it Statistical
Physics} (Pergamon, New York, 1969).
\item{[14]} M. Gyulassy, K. Kajantie, H. Kurki-Suonio and L. McLerran,
Nucl. Phys. B 237 (1984) 477 ; H. Kurki-Suonio, Nucl. Phys. B 255
 (1985) 231.
\item{[15]} R. V. Gavai, TIFR preprint TIFR/TH/92-12, to be published in
``Quantum Fields on the Computer", Ed. M. Creutz, World Scientific.
\item{[16]} K. Binder, Rep. Prog. Phys. 50 (1987) 783.
\item{[17]}  Y. Iwasaki, K. Kanaya, T. Yoshi\'e, T. Hoshino, T. Shirakawa,
    Y. Oyanagi, S. Ichii and T. Kawai, Phys. Rev. Lett. 67 (1991) 3343.
\item{[18]} J. Potvin and C. Rebbi, in Proceedings of ``Lattice 90",
Tallahassee, Eds. U. M. Heller, A. D. Kennedy and S. Sanielevici, Nucl.
Phys. B(PS)20, 309 (1991).
\item{[19]} M. Fukugita, M. Okawa and A. Ukawa, Nucl. Phys. B337 (1990)
181.
\item{[20]} F. R. Brown, F. P. Butler, H. Chen, N. H. Christ, Z. Dong, W.
Schaffer, L. I. Unger and A. Vaccarino, Phys. Rev. Lett. 65 (1990) 2491.

\vfil\end